\documentclass[graybox,vecarrow]{svmult}

\usepackage{mathptmx}       
\usepackage{helvet}         
\usepackage{courier}        
\usepackage{type1cm}        
\usepackage{graphicx}        
\usepackage{multicol}        
\usepackage[bottom]{footmisc}

\usepackage{amssymb}
\usepackage{amsmath} 

\newcommand{\bea}{\begin{eqnarray}}
\newcommand{\eea}{\end{eqnarray}}
\newcommand{\be}{\begin{equation}}
\newcommand{\ee}{\end{equation}}
\newcommand{\rr}{\mathbf{r}}

\newcommand{\kk}{\mathbf{k}}

\newcommand{\qq}{\mathbf{q}}

\begin{document}

\title*{Spatial and temporal coherence of a Bose-condensed gas}
\author{Yvan Castin and Alice Sinatra}
\institute{Yvan Castin \at Laboratoire Kastler Brossel, 
Ecole normale sup\'erieure, CNRS and UPMC, Paris (France),
\email{yvan.castin@lkb.ens.fr}
\and
Alice Sinatra \at Laboratoire Kastler Brossel, 
Ecole normale sup\'erieure, CNRS and UPMC, Paris (France),
\email{alice.sinatra@lkb.ens.fr}}
%
%
\maketitle

\abstract*{
}

\abstract{
The central problem of this chapter is temporal coherence of a three-dimensional spatially homogeneous Bose-condensed gas, initially 
prepared at finite temperature and then evolving as an isolated interacting system. A first theoretical tool is a
number-conserving Bogoliubov approach that allows to describe the system as a weakly interacting gas of quasi-particles. This approach naturally introduces the phase operator of the condensate: a central actor since loss of temporal coherence is governed by the spreading of the condensate phase-change. A second tool is the set of kinetic equations describing the Beliaev-Landau processes for the quasi-particles. We find that in general the variance of the condensate phase-change at long times $t$ is the sum of a ballistic term $\propto t^2$ and a diffusive term $\propto t$ with temperature and interaction dependent coefficients. In the thermodynamic limit, the diffusion coefficient scales as the inverse of the system volume. The coefficient of $t^2$ scales as the inverse volume squared times the variance of the energy of the system in the initial state and can also be obtained by a quantum ergodic theory (the so-called eigenstate thermalisation hypothesis). 
}

\section{Description of the problem}
\label{sub:intro}

We consider a single-spin state Bose gas prepared in equilibrium. To extract the relevant physics, 
we avoid the complication of harmonic trapping present in real experiments \cite{Cornell0,Ketterle0,Hulet0} and we consider a spatially homogeneous system in a parallelepipedic quantization volume $V$ with periodic boundary conditions. In all the chapter except subsection
\ref{sub:nfluc} the total particle number is fixed and equal to $N$. In all the chapter except in subsection 
\ref{sub:low_d}  the system is three-dimensional. We restrict to the deeply Bose-condensed regime
where the non-condensed fraction is small. This implies that the temperature $T$ is much lower than the critical temperature $T_c$ and that the system is weakly interacting. Interactions between the cold bosons are characterized
by the $s$-wave scattering length $a$, that we take positive for repulsive interactions. 
The microscopic details of the interaction potential are irrelevant here since the interaction range is much smaller than
the typical de Broglie wavelength of the particles.
The weakly interacting regime, in the considered low temperature regime, is then defined by $(\rho a^3)^{1/2} \ll 1$
where $\rho=N/V$ is the mean density.

We assume that the gas is prepared in thermal equilibrium at negative times with some unspecified experimental procedure generally implying a coupling with the outer world. For clarity we consider first that the system is prepared either in the canonical or the microcanonical ensemble, then we apply our theory to a more general ensemble:
a statistical mixture of microcanonical ensembles with weak relative energy fluctuations.
After the preparation phase, at positive times, the system is supposed to be {\sl totally isolated} in its evolution. This implies that the total particle number $N$ and the total energy $E$ are exactly conserved in time evolution.
This assumption is realistic for ultra-cold atom experiments: the atoms are hold in conservative immaterial traps and 
the three-body loss rates are very low in the weak density limit. As we shall see, this has important consequences for the temporal coherence of the gas.

A first property that we discuss in this chapter is the spatial coherence of the gas.
This is determined by the first-order coherence function
\be
g_1(\rr) \equiv \langle \hat{\psi}^\dagger(\rr) \hat{\psi}(\mathbf{0})\rangle
\label{eq:g1}
\ee
where the bosonic field operator $\hat{\psi}(\rr)$ annihilates a particle in position $\rr$. The $g_1$ function
has been measured using atomic interferometric techniques 
\cite{Esslinger_Bloch_Hansch_2000}.
In the thermodynamic limit, $g_1(\rr)$ tends to the condensate density {$\rho_0>0$} at large distances $r$.
One refers to this property as long-range order.

A second, more subtle property, that we discuss in detail is the temporal coherence of the gas.
We define the temporal coherence function of the condensate as
\be
\langle a_0^\dagger(t) a_0(0)\rangle
\label{eq:a0t_a0}
\ee
where $a_0$ is the annihilation operator in the condensate mode that is the plane wave with $\kk=\mathbf{0}$.
Contrarily to the case of $g_1$, here the operators appear in the Heisenberg picture at different times.
The temporal coherence function of the condensate is measurable (as we argue in subsection \ref{sub:measure}) but it
was not measured yet. The closest analog that has been measured is the relative coherence of two 
condensates in different external or internal states at equal times \cite{Cornell,Ketterle}.
The coherence time of the condensate is simply the half width of the temporal coherence function.
Remarkably at zero temperature it was shown that the coherence function does not decay at long times,
it rather oscillates \cite{Beliaev}
\be
\langle a_0^\dagger(t) a_0(0)\rangle \sim \langle \hat{n}_0 \rangle e^{i\mu(T=0)t/\hbar}
\ee
where $\langle  \hat{n}_0 \rangle$ is the mean number of particles in the condensate and $\mu(T=0)$ is the 
ground state chemical potential of the gas. This implies an infinite coherence time.
At finite temperature however one expects a finite coherence time for a finite size system.
We find that in the thermodynamic limit this coherence time diverges with a scaling with the system volume $V$ 
that depends on the statistical ensemble in which the system is prepared. 

This chapter is based on our works \cite{Superdiff,Genuine,Truediff}\footnote{Particle losses are not discussed in this chapter. Their effect on temporal coherence is weak at relevant times as explicitly shown in \cite{Truediff} for one-body losses in the canonical ensemble.}. It is 
organized as follows. We give a pedagogical presentation of the number conserving Bogoliubov
theory, a central tool for our problem, in section \ref{sec:Bogol}. We apply this theory to the spatial coherence
in section \ref{sec:spatial}. The more involved issue of temporal coherence is treated in section \ref{sec:temporal}.
In subsection \ref{sub:measure} we discuss how to measure $\langle a_0^\dagger(t) a_0(0)\rangle$ with cold atoms.
General considerations are given in \ref{sub:gen}, showing the central role of condensate phase-change spreading, that is then studied for different
initial states of the gas. First for a single-mode model in \ref{sub:nfluc} and for the canonical ensemble \ref{sub:can}, where one of the conserved quantities (the particle number $N$ or the energy $E$) has fluctuations in the initial state.
Then for the microcanonical ensemble  \ref{sub:micro}, where none of these conserved quantities
fluctuates. Finally in the already mentioned more general statistical ensemble
within a unified theoretical framework  in subsection \ref{sub:summary}.

\section{Reminder of Bogoliubov theory}
\label{sec:Bogol}

The central result of Bogoliubov theory \cite{Bogoliubov} is that our system can be described as an ensemble of weakly interacting
{\sl quasi-particles}. The necessity to go from a particle to a quasi-particle picture to obtain weakly interacting
objects is due to the presence of the condensate that provides a large bosonic enhancement of particle scattering processes in and out of the condensate mode. In the initial work of Bogoliubov the quasi-particles are non-interacting. 
We will need to include the interactions among quasi-particles that give them a finite lifetime
through the so-called Beliaev-Landau mechanism \cite{Beliaev,Giorgini}. Here we present a powerful formulation of Bogoliubov ideas 
introducing the phase operator $\hat{\theta}$ for the condensate mode \cite{Girardeau}: in addition to making the theory number conserving \cite{CastinDum,Gardiner,LesHouches99}, $\hat{\theta}$ will play a crucial role for the study of temporal coherence.

\subsection{ Lattice model Hamiltonian:}
\label{sub:latticeH}
Commonly a zero range delta potential $V_{12}=g \delta(\rr_1-\rr_2)$ is used to model particle interactions with 
an effective coupling constant 
\be
g=\frac{4\pi\hbar^2 a}{m}
\label{eq:defg}
\ee
(here the $s$-wave scattering length is $a>0$ and $m$ is the mass of a particle). This however does not lead to a mathematically well defined Hamiltonian problem, even for two particles. As explained in \cite{LesHouchesLowD} a convenient way to regularize the theory while keeping the simplicity of contact interactions is to discretize the coordinate space on a cubic lattice with lattice spacing $b$. This automatically introduces a cut-off in momentum space, since single particle wave vectors are restricted to the first Brillouin zone (FBZ) of the lattice $[-\frac{\pi}{b},\frac{\pi}{b})^3$.
Then 
\be
V_{12}=g_0 \frac{\delta_{\rr_1,\rr_2}}{b^3}
\ee 
where now $\delta$ is a discrete Kronecker $\delta$. 
The bare coupling constant $g_0$ is adjusted to reproduce the true $s$-wave scattering length 
on the lattice \cite{LesHouchesLowD},  
\be
g_0 = \frac{g}{1- C a/b}
\label{eq:g0}
\ee
where $C=2.442\,749\ldots$ is a numerical constant \footnote{This results from the formula
$g_0^{-1}=g^{-1} - \int_{\rm FBZ} \frac{d^3k}{(2\pi)^3} \frac{m}{\hbar^2 k^2}$.}.
The Bogoliubov method is applicable when the zero energy scattering problem is treatable in the Born regime
\cite{LiebLiniger} which requires here that $a \ll b$. In this limit $g_0\simeq g$. For the lattice model to well describe
continuous space physics the lattice spacing $b$ should be smaller than the macroscopic length scales $\xi$ and
$\lambda$ of the gas. The healing length $\xi$ is defined as
\be
\frac{\hbar^2}{2m \xi^2} = \rho g
\label{eq:heal}
\ee
and the thermal de Broglie wavelength as 
\be
\lambda^2=\frac{2\pi \hbar^2}{m k_BT}
\label{eq:lambda}
\ee 
Note that in the weakly interacting and degenerate limit one has $\xi \gg a$ and $\lambda \gg a$.

The system Hamiltonian in second quantized form is
\be
\hat{H} = \sum_\rr b^3 \left[\hat{\psi}^\dagger h_0 \hat{\psi} + \frac{g_0}{2} \hat{\psi}^\dagger \hat{\psi}^\dagger \hat{\psi} \hat{\psi}\right]
\label{eq:H}
\ee
where $h_0$ is the one-body hamiltonian reduced here to the kinetic energy term, $h_0=-\frac{\hbar^2}{2m} \Delta_\rr$,
with a discrete laplacian reproducing the free wave dispersion relation
$E_k=\hbar^2 k^2/2m$ when applied over a plane wave. The bosonic field operator $\hat{\psi}(\rr)$ obeys the 
discrete commutation relation
\be
[\hat{\psi}(\rr_1),\hat{\psi}^\dagger(\rr_2)] =  \frac{\delta_{\rr_1,\rr_2}}{b^3} 
\ee

\subsection{Bogoliubov expansion of the Hamiltonian}
\label{sub:HBog}
We split the field operator into the condensate field and the non-condensed field $\hat{\psi}_\perp(\rr)$ orthogonal to the condensate wave function $\phi(\rr)$:
\be
\hat{\psi}(\rr) = \phi(\rr) \hat{a}_0 + \hat{\psi}_\perp(\rr)
\label{eq:split}
\ee
where $\hat{a}_0$ is the annihilation operator of a particle in the condensate mode.
For the homogeneous system that we consider, $\phi(\rr)=1/V^{1/2}$. The main idea of the Bogoliubov approach 
is to use the fact that the non-condensed field is much smaller than the condensate field to expand the Hamiltonian in powers 
of $\hat{\psi}_\perp(\rr)$. This becomes truly operational if one succeeds in eliminating the amplitude $\hat{a}_0$
of the field on the condensate mode. For the modulus of $\hat{a}_0$ we can use the identity
\be
\hat{n}_0= \hat{N}-\hat{N}_\perp
\label{eq:n0}
\ee
with $\hat{N}$ the total particle number operator, $\hat{n}_0=\hat{a}_0^\dagger \hat{a}_0$ the condensate particle 
number operator and $\hat{N}_\perp = \sum_\rr b^3 \hat{\psi}_\perp^\dagger \hat{\psi}_\perp$  the non-condensed 
particle number operator. 
The elimination of the phase of $\hat{a}_0$ at the quantum level is more subtle, and it was not performed in the original work
of Bogoliubov. We introduce the modulus-phase representation \cite{Girardeau}
\be
\hat{a}_0 = e^{i\hat{\theta}} \hat{n}_0^{1/2}
\label{eq:a0_gir}
\ee
with the hermitian phase operator $\hat{\theta}$, conjugate to the condensate particle number: 
\be
[\hat{n}_0,\hat{\theta}] = i
\label{eq:comm_ph_n}
\ee
It is known that the introduction of a phase operator in quantum mechanics is a delicate matter
\cite{RevuePhase}. As we explain below, our formulation is not exact but it is extremely accurate in the present case of a highly populated condensate mode.
As it appears from (\ref{eq:comm_ph_n}), there is a formal analogy with the position operator $\hat{x}$
and the momentum operator $\hat{p}$ of a fictitious particle in one spatial dimension.
For the fictitious particle $\hat{p}$ is the generator of spatial translations so that

\medskip
\centerline{
\begin{tabular}{lcl}
$[\hat{x},\hat{p}] =i\hbar$  & $\Longrightarrow$ & $\displaystyle  e^{i\hat{p}/\hbar} |x\rangle = | x - 1\rangle$ \\
$[\hat{n}_0,\hat{\theta}] = i$ & $\Longrightarrow$ & $\displaystyle e^{i\hat{\theta}} |n_0:\phi\rangle = |n_0-1: \phi\rangle$ 
\end{tabular}}

\medskip
\noindent
where $|x\rangle$ represents the fictitious particle localized in position $x$ and $|n_0:\phi\rangle$ is the Fock state 
with $n_0$ particles in the condensate mode. As a consequence the representation (\ref{eq:a0_gir}) of $\hat{a}_0$
has the correct matrix elements in the Fock basis. The operator $\exp(i\hat{\theta})$ is a respectable unitary operator$\ldots$ except when the condensate mode is empty where one gets the meaningless result:
\be
e^{i\hat{\theta}} |0:\phi\rangle \stackrel{?!}{=} |-1 : \phi\rangle
\ee
This is in practice not an issue if, in the physical state of the system, 
the probability for the condensate mode to be empty is negligible.
For a finite size system the probability  distribution of $n_0$ was calculated using the Bogoliubov approach
and even an exact numerical approach \cite{Cartago,Iacopo_N0}. In the thermodynamic limit we expect that the
probability of having an empty condensate vanishes exponentially with the system size at $T<T_c$.

In order to eliminate the condensate phase we introduce the number conserving operator \cite{CastinDum,Gardiner}
\be
\hat{\Lambda}(\rr) = e^{-i\hat{\theta}} \hat{\psi}_\perp(\rr)
\ee
The success of the elimination procedure is guaranteed since the Hamiltonian conserves the particle number:
Injecting the splitting of the field (\ref{eq:split}) in the Hamiltonian and expanding, generates a series of terms in which
$\hat{a}_0$ appears either with $\hat{a}_0^\dagger$ or with $\hat{\psi}_\perp^\dagger(\rr)$.
Expanding  $\hat{H}$ to second order in  $\hat{\psi}_\perp$ and using (\ref{eq:n0}) we obtain the Bogoliubov Hamiltonian
\be
\fbox{$\displaystyle
 \hat{H}_{\rm Bog}\! =\! \frac{g_0 N^2}{2 V}\! +\! \sum_\rr\! b^3 \left[\hat{\Lambda}^\dagger (h_0-\mu_0)
\hat{\Lambda}
+ \mu_0 \left(\frac{1}{2}\hat{\Lambda}^2 + \frac{1}{2} \hat{\Lambda}^{\dagger 2}
+ { 2} \hat{\Lambda}^\dagger \hat{\Lambda}\right)\right]$
}
\label{eq:HBog}
\ee
We have assumed that the total particle number is fixed and equal to $N$ and we have replaced $\hat{N}$ by $N$.
Still, one obtains a grand canonical ensemble for the non-condensed modes, with a chemical potential
$\mu_0=g_0\rho$. The condensate indeed acts as a reservoir of particles for the non-condensed modes.
The expression $\mu_0=g_0\rho$ is in fact the zeroth order approximation (in the non-condensed fraction) to the gas chemical potential. In what follows we shall take 
\be
\mu_0=g \rho
\ee 
which is consistent with the Bogoliubov theory at this order.
The terms $\hat{\Lambda}^\dagger \hat{\Lambda}$ in (\ref{eq:HBog}) represent elastic interactions
between the condensate and the non-condensed particles. They also appear in the simple Hartree-Fock theory.
The terms $\hat{\Lambda}^{\dagger 2}$ and hermitian conjugate represent inelastic interactions where two 
condensate particles collide and are both scattered into non-condensed modes with opposite momenta. 
They are absent in the Hartree-Fock theory and they play a crucial role in explaining the superfluidity of the gas.

\subsection{An ideal gas of quasi-particles}
To extract the physics contained in the Bogoliubov Hamiltonian one has to identify the eigenmodes of the
system putting the quadratic Hamiltonian in a normal form.
We present here a brief overview, a more detailed discussion was given in \cite{LesHouches99,BlaizotRipka}.
In the Heisenberg picture the equations of motion of the field operators are linear, provided one collects 
$\hat{\Lambda}$ and $\hat{\Lambda}^{\dagger}$ into a single unknown: 
\be
i\hbar \partial_t \left(\begin{array}{c} \hat{\Lambda} \\ \hat{\Lambda}^\dagger \end{array}\right)
=
\begin{pmatrix}
h_0+\mu_0 & \mu_0\\ -\mu_0 & -(h_0+\mu_0) 
\end{pmatrix}
\left(\begin{array}{c}{\hat{\Lambda}} \\ \hat{\Lambda}^\dagger \end{array}\right) \equiv \mathcal{L} 
\left(\begin{array}{c}{\hat \Lambda} \\ {\hat \Lambda}^\dagger \end{array}\right)
\ee

The matrix $\mathcal{L}$ is not hermitian for the usual scalar product, but it is ``hermitian" 
for a modified scalar product of signature $(1,-1)$. It has moreover a symmetry property 
ensuring that its eigenvalues come in pairs $\pm \epsilon_k$.

We now expand the field operators over the eigenvectors of $\mathcal{L}$:
\be
\left(\begin{array}{c}{\hat \Lambda}(\rr) \\ {\hat \Lambda}^\dagger(\rr) \end{array}\right) = 
\sum_{\kk\neq \mathbf{0}} \frac{e^{i\kk\cdot\rr}}{V^{1/2}} \left(\begin{array}{c}U_k\\ V_k\end{array}\right)
\hat{b}_\kk + \frac{e^{-i\kk\cdot\rr}}{V^{1/2}} \left(\begin{array}{c}V_k\\ U_k\end{array}\right)
\hat{b}_{\kk}^\dagger
\label{eq:modal_dec}
\ee
with $U_k^2-V_k^2=1$ (this is the normalization condition for the modified scalar product). An explicit 
calculation gives 
\be
U_k+V_k=\frac{1}{U_k-V_k}=\left(\frac{\hbar^2 k^2/2m}{2\mu_0+\hbar^2 k^2/2m}\right)^{1/4}
\ee
The coefficients $\hat{b}_\kk$ and $\hat{b}_{\kk}^\dagger$ obey the usual bosonic commutation relations
e.g. $[\hat{b}_\kk,\hat{b}_{\kk'}^\dagger]=\delta_{\kk,\kk'}$.
Injecting the modal decomposition (\ref{eq:modal_dec}) in the Bogoliubov Hamiltonian (\ref{eq:HBog})
one obtains a Hamiltonian of non-interacting bosons called quasi-particles: 
\be
\fbox{$\displaystyle
\hat{H}_{\rm Bog} = E_0(N) + \sum_{\kk\neq \mathbf{0}} \epsilon_k \hat{b}_k^\dagger \hat{b}_\kk
\ \ \ \mbox{ with}\ \ \ 
\epsilon_k = \left[\frac{\hbar^2 k^2}{2m} \left(\frac{\hbar^2 k^2}{2m}+2\mu_0\right)\right]^{1/2}
$}
\label{eq:epsk}
\ee
The quantity $E_0(N)$ is the Bogoliubov approximation of the ground state energy. It reads
\be
E_0(N)= \frac{g_0 N^2}{2 V} - \sum_{\kk \neq {\bf 0}} \epsilon_k V_k^2
\label{eq:E0}
\ee 
In the continuous space limit $b/\xi \to 0$, the sum over $\kk$ has an ultraviolet  ($k\to\infty$) divergence.
If one replaces $g_0$ by its expression (\ref{eq:g0}) expanded to first order in $a/b$, 
$g_0 \simeq g(1+ Ca/b)$, this exactly compensates the ultraviolet divergence and one recovers the
Lee-Huang-Yang result
\be
E_0(N)=  \frac{g N^2}{2 V} \left[ 1 + \frac{128}{15 \pi^{1/2}} (\rho a^3)^{1/2} \right]
\ee
The Bogoliubov spectrum $\epsilon_k$ starts linearly at low $k$: the quasi-particles are then phonons.
At high $k$ one recovers the free particle spectrum shifted upwards by $\mu_0$: quasi-particles in this limit 
are just particles.
At thermal equilibrium in the canonical ensemble for the original system the Bogoliubov density operator is
\be
\hat{\sigma}=\frac{1}{Z_{\rm Bog}} e^{-\beta \hat{H}_{\rm Bog}} \ \ \ \mbox{with}\ \ \ \beta=1/k_B T
\label{eq:sigma}
\ee
where $Z_{\rm Bog}$ is the partition function in the Bogoliubov approximation.
This density operator in the canonical ensemble for {\sl particles},
corresponds in fact to a grand canonical ensemble, with zero chemical potential,
for the {\sl quasi-particles} whose number is not conserved.

\section{Spatial coherence}
\label{sec:spatial}

In this section we discuss the spatial coherence properties of a weakly interacting Bose-condensed gas,
using the Bogoliubov theory. As expected one finds long range order in the thermodynamic limit.
To complete the discussion we briefly address the case of a low-dimensional system where long range order 
is in general lost (except for the $2D$ gas at zero temperature) but where the ideas of the Bogoliubov method 
can be adapted for quasi-condensates \cite{LesHouchesLowD,PopovLeLivre}.

\subsection{Non-condensed fraction and  $g_1$ function}

In a spatially homogeneous gas, the non-condensed fraction is the ratio of the non-condensed 
density $\langle \hat{\Lambda}^\dagger\hat{\Lambda} \rangle$ and the total density $\rho$.
Using the modal decomposition (\ref{eq:modal_dec}) and the thermal equilibrium state (\ref{eq:sigma}), one obtains 
in the thermodynamic limit in $3D$:
\be
\frac{\langle \hat{N}_\perp\rangle }{N} = \frac{\langle \hat{\Lambda}^\dagger\hat{\Lambda}\rangle}{\rho} = 
\frac{1}{\rho}\int\frac{d^3k}{(2\pi)^3} \left[\frac{U_k^2 +V_k^2}{e^{\beta \epsilon_k}-1} + V_k^2\right]
\label{eq:noncond}
\ee
This integral has no ultraviolet ($k\to\infty$) divergence since $V_k^2=O(1/k^4)$. One can thus take 
the continuous space limit $b\to 0$ and integrate over the whole Fourier space. The integral has no
infrared ($k\to 0$) divergence either, since $U_k^2, V_k^2 = O (1/k)$.
In order for the Bogoliubov theory to be applicable, the non-condensed fraction should be small.
From the result (\ref{eq:noncond}) one can check that this is indeed the case for the degenerate
$\rho \lambda^3 \gg 1$ and weakly interacting $(\rho a^3)^{1/2}\ll 1$ regime.

The first-order coherence function (\ref{eq:g1}) in the thermodynamic limit is given in the Bogoliubov theory by
\be
g_1^{\rm Bog}(\rr) = 
\rho - \int\frac{d^3k}{(2\pi)^3} (1-\cos\kk\cdot\rr) \left[\frac{U_k^2 +V_k^2}{e^{\beta \epsilon_k}-1} + V_k^2\right]
\label{eq:g1Bog}
\ee
where we used the exact relation $\langle \hat{a}_0^\dagger \hat{\psi}_\perp \rangle=0$. In the large $r$ limit,
the contribution of the oscillating term $\cos\kk\cdot\rr$  vanishes and $g_1$ tends to the condensate density.
This implies that spatial coherence extends over the whole system size.

\subsection{In low dimensions} 
\label{sub:low_d}

In a straightforward generalization of (\ref{eq:noncond}) to low dimensions, 
the non-condensed fraction is infrared divergent in $2D$ for $T>0$, and in $1D$ for all $T$: there is no Bose-Einstein condensate in the thermodynamic limit in agreement with the
Mermin-Wagner-Hohenberg theorem \cite{MerminWagner,Hohenberg}.
Nevertheless, in the weakly interacting and degenerate regime there are weak density fluctuations and
weak phase gradients. This is the so called quasi-condensate regime \cite{PopovLeLivre,Gora}.
The main ideas of the Bogoliubov approach can still be applied after the introduction of a modulus-phase 
representation of the field operator $\hat{\psi}$ in each lattice site \cite{MoraCastin}:
\be
\hat{\psi}(\rr)= e^{i\hat{\theta}(\rr)} \sqrt{\hat{\rho}({\rr})}
\label{eq:psi_mod_phase}
\ee
where $\hat{\rho}(\rr) b^d$ and $\hat{\theta}(\rr)$ are conjugate variables similarly to (\ref{eq:comm_ph_n}) and
$d$ is the spatial dimension.
As we discussed in subsection \ref{sub:HBog} and in \cite{LesHouchesLowD}, the modulus-phase representation of the annihilation operator in a given
field mode is accurate if this mode has a negligible probability to be empty. 
This in particular requires that the mean number of particles per lattice site is large,
$\rho b^d\gg 1$.  In the weakly interacting $\rho \xi^d \gg 1$ and degenerate 
$\rho \lambda^d \gg 1$ regime, one can adjust $b$ to satisfy this condition
while keeping $b \ll \xi, \lambda$ so as to well reproduce the continuous space physics. 
In this regime one also finds that the probability distribution of the number of particles
on a given lattice site is strongly peaked around the mean value $\rho b^d\gg 1$,  with a width
much smaller than the mean value, which legitimates the representation (\ref{eq:psi_mod_phase}).

If one blindly applies the plain Bogoliubov result (\ref{eq:g1Bog}) in the absence of a condensate
\footnote{One may wonder in $2D$ about the value of $\mu_0= g_0 \rho$,
since $g_0$ logarithmically depends on the lattice spacing $b$ \cite{MoraCastin}, and
dimensionality reasons prevent from forming a coupling constant $g$ (such that
$g\rho$ is an energy) from the quantities $\hbar$, $m$ and $a$, where $a$ is now the $2D$
scattering length,  given in \cite{Gora,OlshaniiPricoupenko} . According to \cite{MoraCastin}
one simply has to take for $\mu_0$ the gas chemical potential $\mu(T)$.},
one finds that the first-order coherence function $g_1^{\rm Bog}(\rr)\to -\infty$  at infinity,
logarithmically with $r$ in $2D$ ($T>0$) and in $1D$ ($T=0$), and even linearly in $r$
in $1D$ at $T>0$. One may believe at this stage that $g_1^{\rm Bog}(\rr)$ is simply
meaningless in those cases. The extension of the Bogoliubov theory to quasi-condensates
however produces the remarkable result  \cite{MoraCastin}:
\be
g_1^{\rm QC}(\rr)=\rho \exp\Big[\frac{g_1^{\rm Bog}(\rr)}{\rho}-1\Big]
\label{eq:g1QC}
\ee
The quasi-condensate first-order coherence function $g_1^{\rm QC}(\rr)$  tends to
zero for $r\to \infty$ as a power law in $2D$ ($T>0$) and in $1D$ ($T=0$),  and exponentially
for $T>0$ in $1D$, as expected  \cite{PopovLeLivre}. The gas has then a finite coherence length
$l_c$ (e.g.\ the half-width of $g_1$) much larger than $\xi$ or $\lambda$
in the weakly interacting and degenerate regime.
Over distances $r \ll l_c$, phase fluctuations
are small, and the system gives the illusion of being a condensate: one can linearize the exponential in 
Eq.(\ref{eq:g1QC}), to obtain $g_1^{\rm QC}(\rr) \simeq g_1^{\rm Bog}(\rr)$.
The phase and density fluctuation properties of the quasi-condensates at nonzero
temperature have been studied experimentally with
cold atoms in $1D$ \cite{Walraven,Aspect,Bouchoule} and in $2D$
\cite{Dalibard2D,ChengChin} and confirm the theoretical picture.

\section{Temporal coherence}
\label{sec:temporal}

In this section we discuss the temporal coherence properties of a finite size Bose-condensed gas, defined
by the coherence function $\langle \hat{a}_0^\dagger(t) \hat{a}_0(0) \rangle$ already introduced  in equation (\ref{eq:a0t_a0}). 
Although, strictly speaking, this coherence function was not measured yet with cold atoms, we argue in 
section \ref{sub:measure}  that it is in principle measurable. 
In subsection \ref{sub:gen} we show that the condensate coherence function (\ref{eq:a0t_a0}) can be related to
the condensate phase-change during the time interval $t$. The loss of temporal coherence is thus due to the
spreading in time of this phase-change, which is the quantity that we actually calculate.
Whenever one of the conserved quantities (total particle number $N$ or total energy $E$) fluctuates in the initial state
from one realization to the other, the phase-change spreads ballistically. Once the effect of fluctuations of $N$ is understood (subsection \ref{sub:nfluc}), the more involved effect of energy fluctuations for fixed $N$ can be understood by analogy. The resulting guess for the phase-change spreading can be justified within the quantum ergodic theory
(subsection \ref{sub:can}). The only case in which pure phase diffusion is found is when the conserved quantities 
$N$ and $E$ are fixed, that is in the microcanonical ensemble (subsection \ref{sub:micro}). 
For fixed $N$ and a general statistical ensemble for energy fluctuations, we finally give in subsection \ref{sub:summary}  the expression for the variance of the phase-change in the long time limit, that includes both a ballistic term and a diffusive term.

\subsection{How to measure the temporal coherence function}
\label{sub:measure}
We give here an idea of how to measure the condensate temporal coherence function  
$\langle \hat{a}_0^\dagger(t) \hat{a}_0(0) \rangle$ in a cold atom experiment \cite{Truediff}.
The scheme uses two long-lived atomic internal states $|a\rangle$ and $|b\rangle$ and it is a Ramsey experiment as in
\cite{Cornell}, with the notable difference that the pulses are arbitrarily weak instead 
of being $\pi/2$ pulses. 

The Bose-condensed gas is prepared in equilibrium in the internal state $|a\rangle$ and the state $|b\rangle$ is initially empty. At time zero one applies a very weak electromagnetic pulse, of negligible duration, coherently coupling the two internal states.
After the pulse, the system evolves during a time $t$ in presence of interactions only among atoms in $|a\rangle$:
we assume no interactions between $a$ and $b$ components\footnote{This can be realized experimentally either
using a Feshbach resonance \cite{Oberthaler} or spatially separating the two components 
\cite{Treutlein}.} and negligible interactions within the $b$
component due to the very weak density in that component.
At time $t$ one applies a second pulse of the same amplitude, and one measures the particle number in state $|b\rangle$
in the plane wave $\kk={\bf 0}$.

The scheme can be formalized as follows. The first pulse, at $t=0$, coherently mixes the two bosonic fields
$\hat{\psi}_a$ and $\hat{\psi}_b$ with a real amplitude $\eta$ so that
\bea
\hat{\psi}_a(\rr,0^+) &=& \sqrt{1-\eta^2} \, \hat{\psi}_a(\rr,0^-) + \eta \hat{\psi}_b(\rr,0^-) \label{eq:mixa} \\
\hat{\psi}_b(\rr,0^+) &=& \sqrt{1-\eta^2} \, \hat{\psi}_b(\rr,0^-) - \eta \hat{\psi}_a(\rr,0^-) 
\label{eq:mixb}
\eea
In between time $0^+$ and time $t^-$ the two fields evolve independently. Field $\hat{\psi}_a$ evolves in presence of
kinetic and interaction terms as in (\ref{eq:H}). Field $\hat{\psi}_b$ evolves with kinetic and internal energy terms so that
its amplitude on the $\kk={\bf 0}$ mode obeys
\be
\hat{b}_0(t^-) = e^{i \delta t} \hat{b}_0(0^+)
\label{eq:b0_tm}
\ee
where $\delta$ is the detuning between the electromagnetic field and the $a-b$ atomic transition
(the calculation is performed in the rotating frame).
The second pulse at time $t$ mixes again the two fields with the same mixing amplitudes as in 
(\ref{eq:mixa}), (\ref{eq:mixb}).
 After the second pulse one measures $N_{b 0}(t) = \langle (\hat{b}_0^\dagger  \hat{b}_0)(t^+) \rangle$.
Using the mixing relations and (\ref{eq:b0_tm}) one expresses $\hat{b}_0(t^+)$ as a function of 
$\hat{b}_0(0^-),\hat{a}_0(0^-)$ and $\hat{a}_0(t^-)$. Since the initial state for component $b$ is the vacuum, the contribution of $\hat{b}_0(0^-)$ vanishes and
one obtains the exact relation:
\bea
N_{b 0}(t) &=& \eta^2 \left \{ (1-\eta^2) \langle (\hat{a}_0^\dagger  \hat{a}_0)(0^-) \rangle + 
 \langle (\hat{a}_0^\dagger  \hat{a}_0)(t^-) \rangle_{\rm pulse}  \nonumber \right. \\  &+&  \left.
 \sqrt{1-\eta^2} \left[  e^{i\delta t }  \langle \hat{a}_0^\dagger(t^-)  \hat{a}_0(0^-) \rangle_{\rm pulse} + {\rm c.c.}\right]
 \right \}
\eea
that we expand for vanishing $\eta$:
\be
N_{b 0}(t) = 2\eta^2 \left \{  \langle \hat{n}_0 \rangle  
+ {\rm Re}\, \left[  e^{i\delta t}  \langle \hat{a}_0^\dagger(t)  \hat{a}_0(0) \rangle \right] \right \}
+ O(\eta^4)
\ee
In particular, the subscript $\langle \ldots\rangle_{\rm pulse}$ on the expectation values, indicating that 
they are taken for a system having experienced the first pulse, was removed\footnote{The expectation values $\langle \ldots\rangle_{\rm pulse}$ differ from the original ones
$\langle\ldots\rangle$ in the absence of pulse by $O(\eta^2)$: To first order in $\eta$, the perturbation of $\hat{\psi}_a$
due to the pulse 
is linear in $\hat{\psi}_b(0^-)$ and has a zero contribution to the expectation values since component $b$ is initially in vacuum.}.
The desired correlation function $\langle \hat{a}_0^\dagger(t)  \hat{a}_0(0) \rangle$ 
can be extracted from the contrast of the fringes obtained 
by varying the electromagnetic field frequency.
The signal $N_{b 0}(t)$ itself is small (it is proportional to $\eta^2$) but the contrast of the fringes is independent of $\eta$
in the small $\eta$ limit, and it starts at unity at $t=0$. 

\subsection{General considerations about $\langle \hat{a}_0^\dagger(t)  \hat{a}_0(0) \rangle$}
\label{sub:gen}

\subsubsection{Phase-change spreading}
Here we go through a sequence of transformations that relates the temporal coherence function 
$\langle \hat{a}_0^\dagger(t) \hat{a}_0(0) \rangle$ to the variance of the condensate phase-change
$\hat{\theta}(t)-\hat{\theta}(0)$. 
We use the modulus-phase representation (\ref{eq:a0_gir}) of the annihilation operator $\hat{a}_0$. 
Since the non-condensed fraction is very small, we simply neglect the fluctuations of the modulus of $\hat{a}_0$
i.e. we replace $\hat{n}_0$ with its mean value in equation (\ref{eq:a0_gir}). We then obtain 
\footnote{Here we have neglected the non-commutation of $\hat{\theta}(t)$ and $\hat{\theta}(0)$. 
From the Baker-Campbell-Hausdorff formula, and to zeroth order in the non-condensed fraction, see equation (\ref{eq:nfluc_theta_dot}), the correction is a factor $e^{-\frac{it}{2\hbar}\mu'(N)+O(N^{-2})}$ which is irrelevant for our discussion.}
\be
\langle \hat{a}_0^\dagger(t) \hat{a}_0(0)\rangle \simeq \langle \hat{n}_0\rangle \langle e^{-i[\hat{\theta}(t)-\hat{\theta}(0)]}\rangle
\ee
If the phase-change { $\hat{\theta}(t)-\hat{\theta}(0)$} has a Gaussian distribution, which may be checked {\it a posteriori},
the application of Wick's theorem yields
\be
\fbox{$\displaystyle
\langle \hat{a}_0^\dagger(t) \hat{a}_0(0)\rangle \simeq \langle \hat{n}_0\rangle e^{-i \langle \hat{\theta}(t)- \hat{\theta}(0)\rangle}
e^{-{\rm Var}\,[\hat{\theta}(t)-\hat{\theta}(0)]/2}$}
\label{eq:ifGauss}
\ee
This remarkable formula quantitatively relates the loss of temporal coherence in an isolated Bose-condensed gas to the spreading of the condensate phase-change. 

\begin{figure}[htb]
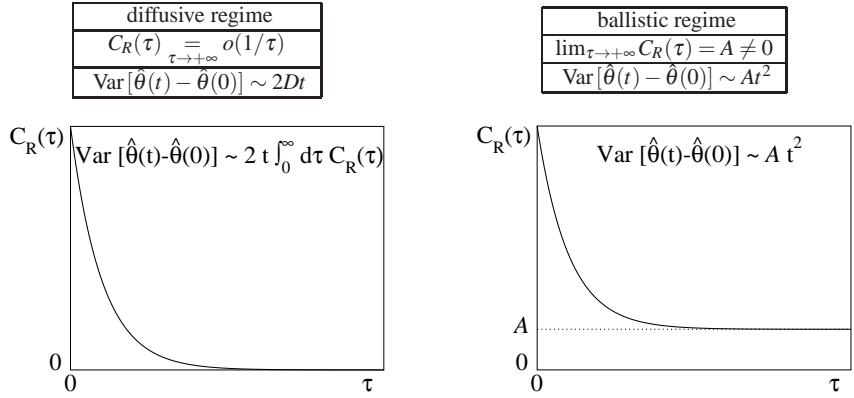

\begin{center}
\begin{tabular}{ccc}
\begin{tabular}{|c|}
\hline
 { diffusive regime} \\
\hline
 $C_R(\tau)\underset{\tau\to+\infty}=o(1/\tau)$ \\
\hline
 \: $\mbox{ Var}\,[\hat{\theta}(t)-\hat{\theta}(0)]\sim 2D t$ \:  \\
\hline
\end{tabular} & &
\begin{tabular}{|c|}
\hline
{ ballistic regime}  \\
\hline
$\lim_{\tau\to+\infty} C_R(\tau)= A \neq 0$ \\
\hline
 \: $\mbox{ Var}\,[\hat{\theta}(t)-\hat{\theta}(0)]\sim A t^2$  \:  \\
\hline
\end{tabular} \\ & \\
\includegraphics[width=5cm,clip=]{figca.eps}& \hspace{1cm} &\includegraphics[width=5cm,clip=]{figcb.eps} 
\end{tabular}
\caption{Different regimes of the condensate phase-change spreading at long times. $C_R$ is the real part of the correlation function $C$ defined in (\ref{eq:C}).
\label{fig:C}}
\end{center}
\end{figure}

The operational way to determine the condensate phase-change spreading is to work with the phase derivative:
contrarily to $\hat{\theta}$, $\dot{\hat \theta}$ is a {\it single-valued} hermitian operator that has a simple expression within
the Bogoliubov approach. The correlation function of the phase derivative 
\be
C(t)=\langle \dot{\hat{\theta}}(t) \dot{\hat{\theta}}(0)\rangle-\langle \dot{\hat \theta}\rangle^2
\label{eq:C}
\ee
gives access to the variance of the phase-change by simple integration:
\be
\fbox{$\displaystyle
\mbox{ Var}\,[\hat{\theta}(t)-\hat{\theta}(0)] = 2t\, \int_0^t d\tau\, C_R(\tau) - 2\, \int_0^t d\tau\, \tau \, C_R(\tau)
$}
\ee
where $C_R$ is the real part of $C$.
One obtains a single integral (rather than a double integral) using the fact that the real part of
$\langle \dot{\hat{\theta}}(t_1)\dot{\hat{\theta}}(t_2) \rangle$ is a function of $|t_1-t_2|$ only, for a system at
equilibrium.
The long-time behavior of $C_R$ determines how the phase-change spreads at long times as summarized
in Fig.~\ref{fig:C}. 

At finite temperature, one might expect that $\dot{\hat{\theta}}(t)$ decorrelates 
from $\dot{\hat{\theta}}(0)$ at long times so that $C_R \to 0$ and the phase-change spreading is diffusive. 
As we will see, this is however not the case, except if the system is prepared in the microcanonical ensemble.
This is a consequence of energy conservation between times $0$ and $t$ in our isolated system.
This point was overlooked in the early studies of \cite{ZollerGardiner,Graham1,Graham2} where the non-condensed modes were treated 
as a Markovian reservoir and phase diffusion was predicted. A subsequent study \cite{Kuklov} based on a  many-body Hamiltonian approach  showed that phase-change spreading is ballistic for a system prepared in the canonical ensemble. The coefficient of $t^2$ in \cite{Kuklov} was however calculated within the pure Bogoliubov approximation, neglecting the interactions between the Bogoliubov quasi-particles, which is illegitimate in the long time limit as we shall see. 

\subsubsection{Key ingredients of the theory}
In order to correctly determine the phase-change spreading in the long time limit, we shall use two key ingredients
in our theoretical treatment: an accurate expression of the phase
derivative and the inclusion of the interactions among Bogoliubov quasi-particles, to which we add the constraint of strict energy conservation during the system evolution.

\noindent
{\sl Time derivative of condensate phase operator:}
The commutator of $\hat{\theta}$ with $\hat{H}$ given by (\ref{eq:H}) is calculated exactly using
\be
[\hat{\theta},\hat{\psi}(\rr)] =  - \hat{a}_0  \frac{i \phi(\rr)}{2 \hat{n}_0}
\ee
and its hermitian conjugate, with the condensate wave function $\phi(\rr)=1/V^{1/2}$. The exact result is given 
in equation (67) of \cite{Superdiff}. Expanding up to second order in the non-condensed field $\hat{\Lambda}$ 
and using the modal decomposition (\ref{eq:modal_dec}), one obtains for fixed $N$: \footnote{We have neglected oscillating terms
in $\hat{b}\hat{b}$ and $\hat{b}^\dagger \hat{b}^\dagger$: after time integration of $\dot{\hat \theta}$ they give a negligible contribution to $\hat{\theta}(t)-\hat{\theta}(0)$.}
\be
\fbox{$\displaystyle
\dot{\hat \theta} = \frac{1}{i\hbar} [\hat{\theta},\hat{H}] \simeq 
-\frac{1}{\hbar} \mu(T=0) - \frac{g_0}{\hbar V} \sum_{\kk\neq\mathbf{0}} 
(U_k+V_k)^2 \hat{n}_\kk $}
\label{eq:thdot}
\ee
We have introduced the zero-temperature chemical potential
$\mu(T=0)=\frac{d}{dN} E_0(N)$, where $E_0(N)$ is given in (\ref{eq:E0}), and
the quasi-particle number operators 
\be
\hat{n}_\kk=\hat{b}_\kk^\dagger \hat{b}_\kk
\label{eq:nk}
\ee
The expression (\ref{eq:thdot}) of the phase derivative differs from the one heuristically introduced
in \cite{Graham1,Graham2}: $\dot{\hat \theta}$ is not simply equal to $-g \hat{n}_0/\hbar V$. 

\medskip
\noindent
{\sl Interactions between quasi-particles}:
Pushing one step further the Bogoliubov expansion of section \ref{sec:Bogol}, that is including terms up to
third order in the non-condensed field, one obtains
\be
\hat{H} \simeq \hat{H}_{\rm Bog} + \hat{H}_3
\label{eq:Hcub}
\ee
where $\hat{H}_{\rm Bog}$ is the Bogoliubov Hamiltonian (\ref{eq:epsk}) and
\be
\hat{H}_3 = g_0\rho^{1/2} \sum_\rr b^3 \hat{\Lambda}^+ (\hat{\Lambda}+\hat{\Lambda}^\dagger) \hat{\Lambda}
\label{eq:H3}
\ee
The Hamiltonian $\hat{H}_3$ is cubic in the field $\hat{\Lambda}$ and it corresponds to interactions between 
quasi-particles. While $\hat{H}_{\rm Bog}$ is integrable (all the $\hat{n}_\kk$ are conserved quantities), the Hamiltonian $\hat{H}_{\rm Bog} + \hat{H}_3$  is not integrable, which plays a central role in condensate dephasing.
By replacing  $\hat{\Lambda}$ with its modal decomposition (\ref{eq:modal_dec}) in $\hat{H}_3$, two types of resonant processes appear, that do not conserve the total number of quasi-particles: the $\hat{b}^\dagger \hat{b}^\dagger \hat{b}$ Beliaev process and the 
$\hat{b}^\dagger \hat{b} \hat{b}$ Landau process. In the Beliaev process one quasi-particle decays into two 
quasi-particles, while in the Landau process two quasi-particles merge into another quasi-particle. 
The processes involving $\hat{b}^\dagger \hat{b}^\dagger \hat{b}^\dagger$ and $\hat{b}\hat{b}\hat{b}$ are non-resonant
(they do not conserve the Bogoliubov energy) and they cannot induce real transitions at the present order.

\subsection{If $N$ fluctuates}
\label{sub:nfluc}
In this subsection we allow fluctuations of the total number of particles and we investigate their effect on temporal coherence.
The effect is already present in the case of a pure condensate, so that we restrict to a one-mode model in this
subsection: identifying  the condensate particle number $\hat{n}_0$ with the total particle number $\hat{N}$, we obtain
the model Hamiltonian
\be
\hat{H}_{\rm one\ mode} = \frac{g}{2V} \hat{N}^2
\ee
The condensate phase derivative is
\be
\dot{\hat \theta}(t) = \frac{1}{i\hbar} [\hat{\theta},\hat{H}_{\rm one\ mode}] = -\mu(\hat{N})/\hbar
\label{eq:nfluc_theta_dot}
\ee
where the chemical potential for the system with $N$ particles is simply $\mu(N)=gN/V$ for the one-mode model.
Since $\hat{N}$ is a constant of motion, temporal integration is straightforward:
\be
\hat{\theta}(t)-\hat{\theta}(0) = -\mu(\hat{N}) \, t/\hbar
\label{eq:toto}
\ee
If $N$ is fixed there is no phase-change spreading. If the initial state is prepared with fluctuations in $N$ then
the phase-change spreads ballistically \cite{Sols1994,Lewenstein1996}: 
\be
\mbox{ Var}\,[\hat{\theta}(t)-\hat{\theta}(0)] = (t/\hbar)^2  \left(\frac{d\mu}{dN}\right)^2 \, \mbox{ Var}\, \hat{N}
\ee
Correspondingly the temporal coherence function $\langle \hat{a}_0^\dagger(t) \hat{a}_0\rangle$ decays as a Gaussian in 
time\footnote{The phase revivals at macroscopic times multiples of $2 \pi \hbar V/g$ \cite{Walls1996,CastinDalibard1997} 
are absent here due to the Gaussian hypothesis used to obtain (\ref{eq:ifGauss}).}\cite{Walls1996,CastinDalibard1997}.
A similar phenomenon was observed 
experimentally \cite{BlochHansch2002,PritchardKetterle2006,Reichel2010} not for the temporal correlation of a single 
condensate but for equal-time coherence  $\langle \hat{a}_0^\dagger(t) \hat{b}_0(t)\rangle$ between two condensates prepared in different modes or internal states with a well defined relative phase and fluctuations in the relative particle number.

\subsection{$N$ fixed, $E$ fluctuates: Canonical ensemble}
\label{sub:can}

We assume in this subsection that the gas is prepared in equilibrium at finite temperature $T$ in the canonical
ensemble with $N$ particles.
We first treat this case by analogy with the previous subsection, and then we expose a systematic derivation
of the result based on quantum ergodicity.

\subsubsection{Using an analogy with the case of fluctuating $\hat{N}$}
Similarly to $\hat{N}$ in the previous subsection, here $\hat{H}$ is a conserved quantity that fluctuates in the initial state.
Indeed the canonical ensemble is a statistical mixture of energy eigenstates with different eigenenergies. By analogy with (\ref{eq:toto}) we 
expect that
\be
\hat{\theta}(t)-\hat{\theta}(0) \sim -\mu_{\rm mc}(\hat{H})\, t/\hbar
\label{eq:analogy}
\ee
where $\mu_{\rm mc}(E)$ is the chemical potential of the microcanonical ensemble of energy $E$.
As relative energy fluctuations are vanishingly small for a large system, we can linearize $\mu_{\rm mc}(E)$
around the mean energy $\bar E$ to obtain a {\sl ballistic} phase-change spreading
\be
\mbox{ Var}\,[\hat{\theta}(t)-\hat{\theta}(0)]\sim (t/\hbar)^2 \left[\frac{d\mu_{\rm mc}}{dE}(\bar{E})\right]^2\, \mbox{ Var}\, 
\hat{H}
\label{eq:res_bal}
\ee
The coefficient of $t^2$ is proportional to the variance of the energy in the initial state and scales as the inverse
of the system volume in the thermodynamic limit. For convenience, one can
reexpress this coefficient in terms of canonical ensemble quantities using 
$\mu_{\rm mc}[{\bar E}(T)] = \mu(T) $ (for a large system) so that
$\frac{d}{dE} \mu_{\rm mc}({\bar E}) = \frac{d}{dT} \mu /  \frac{d}{dT} {\bar E}$, 
where $\mu(T)$ and $\bar{E}(T)$ are the chemical potential and mean energy in the canonical ensemble at temperature $T$.
An explicit expression of the coefficient of $t^2$ is given in Eq.~(73) of \cite{Superdiff} using Bogoliubov theory to evaluate the partition function, $ {\bar E}(T)$ and $\mu(T)$. The obtained formula for $\mu(T)$ also gives the intuitive and interesting side result
\be
\langle \dot{\hat\theta}\rangle=-\mu(T)/\hbar
\label{eq:thdotmoy}
\ee

\subsubsection{From quantum ergodic theory}

In the previous analogy leading to (\ref{eq:res_bal}) there is a strong implicit hypothesis. The fact that the 
phase-change is a function of the Hamiltonian only, see Eq.~(\ref{eq:analogy}),  is in general true only for an ergodic system in the long time limit. For example if the Hamiltonian was truly equal to $\hat{H}_{\rm Bog}$, $\hat{\theta}(t)-\hat{\theta}(0)$
would depend on the set of all occupation number operators $\hat{n}_\kk$ and Eqs.~(\ref{eq:analogy},\ref{eq:res_bal}) would not apply.

We now derive Eq.~(\ref{eq:res_bal}) using quantum ergodic theory. To this end we calculate the asymptotic value
of the correlation function $C(t)$. To eliminate oscillations of $C(t)$ we evaluate its time average.
By inserting a closure relation over exact eigenstates $|\Psi_\lambda \rangle$ with eigenenergies $E_\lambda$
of the interacting many-body system, we obtain
\be
\frac{1}{t} \int_0^t \, d\tau \; C(\tau) \; \underset{t \to \infty}{\to} \; \sum_\lambda p_\lambda 
|\langle \Psi_\lambda |\dot{\hat{\theta}}| \Psi_\lambda \rangle|^2 -
\left( \sum_\lambda p_\lambda 
\langle \Psi_\lambda |\dot{\hat{\theta}}| \Psi_\lambda \rangle \right)^2
\label{eq:Cergo}
\ee
where $p_\lambda$ is the probability to find the system in the eigenstate $|\Psi_\lambda \rangle$. In the canonical 
ensemble $p_\lambda=\exp(-\beta E_\lambda)/Z$. In (\ref{eq:Cergo}) we have assumed that there are no degeneracies
consistently with the non-integrability of the system\footnote{For a large system the level-spacing $\delta E$ vanishes exponentially with the system size, and one may fear that an exponentially long time $t>\hbar/\delta E$ 
is needed to reach the limit (\ref{eq:Cergo}). However, the corresponding off-diagonal matrix elements of $\dot{\hat{\theta}}$ also vanish exponentially with the system size in the
eigenstate thermalization hypothesis \cite{Rigol}.}.
For a classical system, ergodicity implies that the time average over a trajectory of energy $E$ coincides with the microcanonical average at that energy. The extension of this concept to a quantum system is the so-called
eigenstate thermalization hypothesis \cite{Rigol,Deutsch1991,Olshanii}: the mean value of a few-body observable 
$\hat{O}$
in a {\sl single} eigenstate $|\Psi_\lambda \rangle$ is very close to the microcanonical average at the same energy:
\be
\langle \Psi_\lambda | \hat{O} |\Psi_\lambda\rangle \simeq \bar{\hat O}_{\rm mc}(E=E_\lambda)
\ee
We apply this hypothesis to the operator $\hat{O}=\dot{\hat \theta}$. The last step 
is to realize that within the Bogoliubov theory, the microcanonical average of $\dot{\hat \theta}$
is proportional to the microcanonical chemical potential \footnote{See reference [45] of \cite{Superdiff}.
In fact for a large system it is sufficient to prove the equality in the canonical ensemble of mean energy $E$,
as already given by Eq.~(\ref{eq:thdotmoy}).
}
\be
\bar{\dot{\hat \theta}}\,_{\rm mc}(E)=-\mu_{\rm mc}(E)/\hbar
\label{eq:thetadot_mc}
\ee
One then obtains
\be
\fbox{$\displaystyle
\mbox{Var} [\hat{\theta}(t)-\hat{\theta}(0)] \underset{t \to \infty}{\sim}  \frac{t^2}{\hbar^2} \mbox{Var} \, \mu_{\rm mc}(\hat{H})
$}
\label{eq:res_erg2}
\ee
Linearizing $ \mu_{\rm mc}(\hat{H})$ in (\ref{eq:res_erg2}) for small relative energy fluctuations around ${\bar E}$
one recovers (\ref{eq:res_bal}).

\subsubsection{Physical implications}

A consequence of (\ref{eq:res_bal}) is that, for a system prepared in the canonical ensemble, the correlation function $C(\tau)$ of $\dot{\theta}$ {\sl does not} tend to zero when $\tau \to +\infty$. The same conclusion is reached for
the correlation function of $\hat{n}_0$, whose long time limit can be calculated with the quantum ergodic 
theory \cite{Superdiff}.
This qualitatively contradicts \cite{ZollerGardiner,Graham1,Graham2}. 
It only qualitatively agrees with \cite{Kuklov}
since the system Hamiltonian $\hat{H}$ in \cite{Kuklov} was eventually replaced by the integrable Hamiltonian
$\hat{H}_{\rm Bog}$.

In \cite{ZollerGardiner,Graham1,Graham2} the non-condensed modes were treated as a Markovian reservoir. 
This approximation is excellent to calculate temporal correlation functions of ``microscopic" observables such as the quasiparticle numbers. For example, this gives for $\kk,\kk' \neq {\bf 0}$ \cite{Superdiff}:  
\be
\langle \hat{n}_\kk(t) \hat{n}_{\kk'}(0) \rangle - \langle \hat{n}_\kk \rangle \langle \hat{n}_{\kk'} \rangle
\stackrel{\rm Markov}{=} \delta_{\kk,\kk'}
\langle \hat{n}_\kk \rangle (1+ \langle \hat{n}_\kk \rangle) e^{-\Gamma_\kk t}
\label{eq:Gauss}
\ee
where the damping rate $\Gamma_\kk$ is due to the Beliaev-Landau processes. 
However quantum ergodic theory shows that the exact long time limit of this correlation function is nonzero 
(even for $\kk \neq \kk'$) but rather a quantity of order $1/N$. In the double sum over $\kk$ and $\kk'$ that 
appears in $C(\tau)$, this introduces a macroscopic correction of order $N$ missed by the Markovian
approximation. 

\begin{figure}[t]\sidecaption
{\includegraphics[width=7cm,clip=]{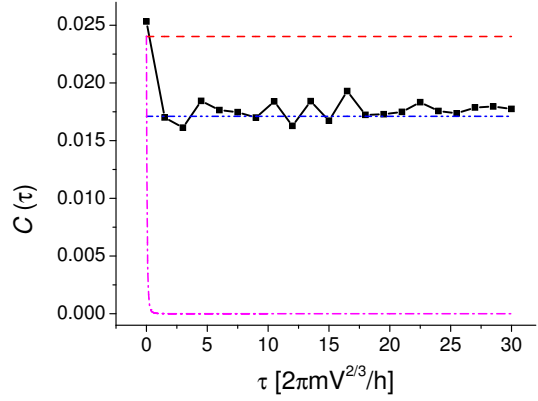}}
\caption{For a gas prepared in the canonical ensemble, correlation function of $\dot{\hat \theta}$ for the classical field model.
The equation of motion is the non-linear Schr\"odinger equation. This corresponds to Fig.~6 in
\cite{Superdiff}. $V$ is the volume. See text for the meaning of the various curves and symbols.
\label{fig:class_superdiff}}
\end{figure}

We illustrate this discussion in Fig.\ref{fig:class_superdiff} with a classical field model \cite{Superdiff}.
The exact numerical result (black squares linked by a solid line) confirms the ergodic result 
(dash-dot-dotted blue curve). The flat red dashed line is the Bogoliubov theory where the $n_\kk$ are constants of
motion. It is close to the numerical result only at short times. The dash-dotted violet curve that tends rapidly to zero
is a Markovian model based on (\ref{eq:Gauss}).

\subsection{$N$ fixed, $E$ fixed: Microcanonical ensemble}
\label{sub:micro}

In this section we assume that the gas is prepared in the microcanonical ensemble of energy $E$.
According to (\ref{eq:res_erg2}) the coefficient of the ballistic spreading of the phase-change is zero. 
It was found in \cite{Truediff} that $C(\tau) = O(1/\tau^3)$ at long times,
so that the phase-change spreads diffusively, with a diffusion coefficient defined by
\be
\mbox{Var}\,[\hat{\theta}(t)-\hat{\theta}(0)] \sim 2 D t \hspace{0.5cm} \mbox{with} \hspace{0.5cm} D=\int_0^\infty d\tau \, C_R(\tau) 
\ee
To determine $D$ we thus need the whole time dependence of $C(\tau)$. From (\ref{eq:thdot}), 
$C(\tau)$ can be deduced from all the correlation functions $\langle \hat{n}_\kk(\tau)\hat{n}_{\kk'}(0)\rangle$
of the quasi-particle number operators.
Within the Bogoliubov approximation for the initial equilibrium state, the gas is prepared in a 
statistical mixture of Fock states $|\{n_\qq^0\}\rangle$ of {\sl quasi-particles} 
 where, in any given Bogoliubov mode of wave vector $\qq$, 
there are exactly $n_\qq^0$ quasi-particles ($n_\qq^0$ is an integer). 
One can then calculate the correlation functions for an initial Fock state $|\{n_\qq^0\}\rangle$ and average 
over the microcanonical probability distribution for the $\{n_\qq^0\}$.

For a given initial Fock state, one then simply needs 
\be
{n}_\kk(\tau) \equiv \langle \{n_\qq^0\}|\hat{n}_\kk(\tau)|\{n_\qq^0\}\rangle
\ee
In the thermodynamic limit,
the evolution of such mean numbers of quasi-particles are given by quantum kinetic equations including
the Beliaev-Landau processes due to $\hat{H}_3$ \cite{Landau}:
\bea
\dot{n}_{{\bf k}} = -\frac{g^2 \rho}{\hbar \pi^2} \int d^3q
\left[ n_{{\bf k}} n_{{\bf q}}- n_{{\bf k}+{\bf q}} (1+ n_{{\bf q}} +
n_{{\bf k}})\right]
\left({\cal A}_{q,k}^{|{\bf k}+{\bf q}|} \right)^2 
 \delta(\epsilon_k+\epsilon_q-\epsilon_{|{\bf k}+{\bf q}|})  && \nonumber \\
\! \! \! -\frac{g^2 \rho}{2\hbar \pi^2} \int d^3q
\left[ n_{{\bf k}}(1+n_{{\bf q}}+n_{{\bf k}-{\bf q}}) - n_{{\bf q}} n_{{\bf k}-{\bf q}} \right]
\left({\cal A}_{q,|{\bf k}-{\bf q}|}^{k} \right)^2 
\delta(\epsilon_q+\epsilon_{|{\bf k}-{\bf q}|}-\epsilon_k)  &&
\label{eq:cin_eqs}
\eea
with the Beliaev-Landau coupling amplitudes:
\be
{\cal A}_{k,k'}^q = U_q U_{k} U_{k'} + V_q V_{k} V_{k'}
+ (U_q+V_q)(V_{k} U_{k'} +U_{k} V_{k'}) 
\ee
The first line in (\ref{eq:cin_eqs}) describes Landau processes and the second line describes Beliaev processes.
In practice we linearize the kinetic equation (\ref{eq:cin_eqs}) around the equilibrium solution $\bar{n}_\kk$
\footnote{For an infinite system, the stationary solution of (\ref{eq:cin_eqs})
is ensemble independent and corresponds to the Bose formula 
$\bar{n}_\kk(E) =1/(\exp{\beta \epsilon_k}-1)$, where $\beta$ is adjusted to give
the mean energy $E$. Finite size effects on the $\bar{n}_\kk$, that can be calculated from 
Eq.~(61) of \cite{Superdiff}, are here not relevant.}
and we solve the resulting linear system numerically. We refer to \cite{Truediff} for technical details.

\begin{figure}[b]\sidecaption
{\includegraphics[width=6cm,clip=]{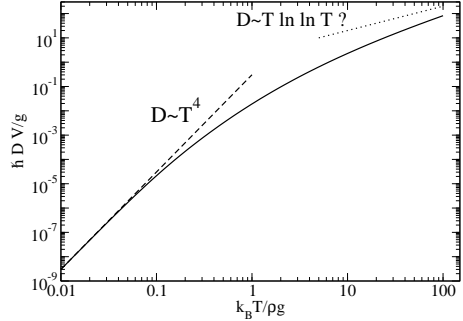}}
\caption{Solid line: universal result for the phase diffusion coefficient in the Bogoliubov limit $(\rho a^3)^{1/2}\ll 1$, $T\ll T_c$.
Dashed line: low-$T$ analytical result (\ref{eq:DlowT}). The high-$T$ behavior is only conjectured, and the dotted line
is an arbitrary linear function of $T$ to guide the eye. $V$ is the volume and $g$ the effective coupling constant 
(\ref{eq:defg}).
\label{fig:diff}}
\end{figure}

The phase diffusion coefficient is shown in Fig.\ref{fig:diff} as a function of the temperature $T$ such that 
the mean canonical energy $\bar{E}(T)$ is equal to the microcanonical energy $E$.
Remarkably, when $D$ and $T$ are properly rescaled (as in the figure), the curve is universal. In particular
this shows that $D$ vanishes as the inverse of the system volume in the thermodynamic limit.
Interestingly, at low temperature, $D$ vanishes with the same power-law $T^4$ as the normal fraction of the gas:
\be
\frac{\hbar D V}{g} \sim 0.3036 \left( \frac{k_BT}{\rho g} \right )^4
\label{eq:DlowT}
\ee

\begin{figure}[b]
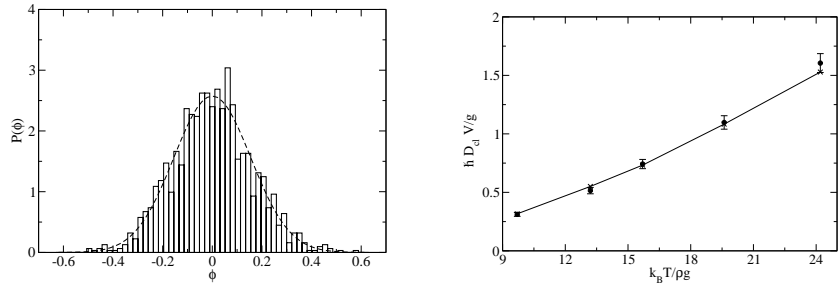

\centerline{\includegraphics[width=5cm,clip=]{classmca.eps}\hspace{1cm}\includegraphics[width=5cm,clip=]{classmcb.eps}}
\caption{Classical field simulations in the microcanonical ensemble. Left panel (taken from \cite{Genuine}): Probability distribution $P(\phi)$ of the condensate phase-change fluctuations $\phi=\theta(t)-\theta(0) - 
\langle \theta(t)- \theta(0) \rangle$ at a large time $t$. The dashed line is the expected Gaussian. Right panel (taken from \cite{Truediff}): Diffusion coefficient as a function of the temperature, extracted from the numerics (bullets with error bars) and calculated by the classical field version of the kinetic equations (\ref{eq:cin_eqs}) (crosses linked by segments). 
\label{fig:Gauss}}
\end{figure}

We performed classical field simulations in the microcanonical ensemble \cite{Genuine}. As expected we found 
that the phase-change has a diffusive behavior: its variance increases linearly in time at long times (not shown) and the phase-change probability distribution is well adjusted by a Gaussian as we show in the left panel of Fig.~\ref{fig:Gauss}.
In the right panel Fig.~\ref{fig:Gauss} we show that the diffusion coefficient is well reproduced by a classical
field version of the kinetic theory.

\subsection{A general statistical ensemble}
\label{sub:summary}

We now consider a generalized ensemble at fixed $N$ that includes both the microcanonical and the canonical ensembles as particular cases.  This is a statistical mixture of microcanonical ensembles with a probability distribution $P(E)$ of the 
system energy $E$ that depends on the particular experimental procedure to prepare the initial state of the gas.
Remarkably the approach of the previous subsection based on kinetic equations can be extended to this case.

\subsubsection{General result for the phase-change spreading}
\label{subsub:general}

Provided that the relative energy fluctuations vanish in the thermodynamic limit, we find the long time limit \cite{Truediff}
\be
\fbox{$\displaystyle
\mbox{Var}\,[\hat{\theta}(t)-\hat{\theta}(0)] \underset{t\to+\infty}{=} \mbox{ Var}\,(E)\, 
\left[\frac{d\mu_{\rm mc}}{\hbar \, dE}(\bar{E})\right]^2 t^2
+ 2 D ( t-t_{\rm off}) +O\left(\frac{1}{t}\right)$}
\label{eq:gen_res}
\ee
For the coefficient $A$ of the ballistic $t^2$ term we recover the {\sl form} of the quantum ergodic result (\ref{eq:res_bal}). 
This is not surprising as the reasoning of subsection \ref{sub:can} does not rely on the fact that the system is prepared in 
the canonical ensemble. On the other hand the {\sl value} of the coefficient does depend on the statistical ensemble
through the mean energy ${\bar E}$ and the variance of the energy. A physical derivation of this result within kinetic theory is given in the next subsection.

A remarkable result is that, in the general ensemble, the phase derivative correlation function $C(\tau)$ is the sum of its long time limit $A$ and of the correlation function $C_{\rm mc}(\tau)$ in the microcanonical ensemble of energy ${\bar E}$:
\be
C(\tau) = A + C_{\rm mc}(\tau)
\ee
As a consequence the diffusion coefficient $D$ of Eq.~(\ref{eq:gen_res})
is the same as the one for the microcanonical ensemble of energy $\bar{E}$. The same conclusion holds for the
constant time offset $t_{\rm off}$ \footnote{This is true to leading order in the system size since our linearized kinetic approach cannot access the subleading terms.}:
\bea
D &=& \int_0^\infty \,d\tau \, C_{R,{\rm mc}}(\tau)  \\
t_{\rm off} &=&  \frac{\int_0^\infty \,d\tau \,\tau \, C_{R,{\rm mc}}(\tau) }{ \int_0^\infty \,d\tau \, C_{R,{\rm mc}}(\tau)}
\label{eq:toff}
\eea
where $C_{R,{\rm mc}}$ is the real part of $C_{\rm mc}$.
The physical origin of the time offset $t_{\rm off}$ is apparent in Eq.(\ref{eq:toff}): it is due to the finite width of the 
phase derivative correlation function.  As $C_{R,{\rm mc}}(\tau)$ is found to be positive, $t_{\rm off}$ can be simply interpreted as the correlation time of the phase derivative in the microcanonical ensemble.
The formal expressions for $D$ and $t_{\rm off}$, in terms of the matrix of the linearized kinetic equations, are given in \cite{Truediff}. 

\begin{figure}[h]
\centerline{\includegraphics[width=8cm,clip=]{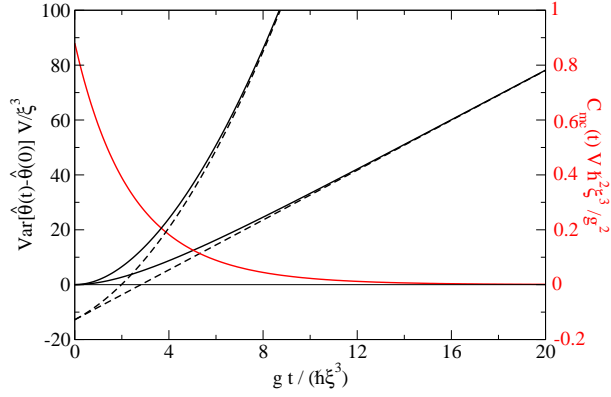}}
\caption{For a quantum system in the thermodynamic limit, the microcanonical phase derivative correlation function 
$C_{\rm mc}(t)$ (red solid line, right vertical axis) and the variance of the phase-change (black lines, left vertical axis) are shown as functions of time. For the variance, the upper (lower) solid line is for the canonical (microcanonical) ensemble, and the dashed lines are the corresponding asymptotic forms of Eq.~(\ref{eq:gen_res}). $k_BT=10 \rho g$,
$V$ is the system volume, $g$ is the effective coupling constant (\ref{eq:defg}) and $\xi$ is the healing length (\ref{eq:heal}). This is Fig.~3 of \cite{Truediff}. In atomic condensates $\xi$ is in the $\mu {\rm m}$ range and the time
unit of the figure is in the ms range. 
\label{fig:cvar}}
\end{figure}

These results are made more concrete by Fig.~\ref{fig:cvar}: for a quantum system
in the thermodynamic limit, we show the microcanonical correlation function
$C_{\rm mc}(t)$ as a function of time, and the variance of the phase-change either in the canonical ensemble of temperature $k_B T=10 \rho g$ or in the microcanonical ensemble  with the same mean energy. 
This reveals in particular that the asymptotic expression (\ref{eq:gen_res}) becomes rapidly accurate.

\subsubsection{Recovering the ballistic spreading from kinetic theory}
Due to energy conservation, the linearized kinetic equations have a zero-frequency undamped mode. 
We will show that, in presence of energy fluctuations in the initial state, the amplitude over this mode is nonzero,
so that the phase derivative correlation function $C(\tau)$ does not tend to zero at long times and the phase-change variance shows a $t^2$ term as in Eq.~(\ref{eq:gen_res}). The derivation presented here was significantly
simplified with respect to the original one of \cite{Truediff}.

We introduce the notation
\be
\bar{n}_\kk(E) = \bar{\hat{n}}_{\kk \,{\rm mc}} (E)
\ee
for the average number of quasi-particles in mode $\kk$ in the microcanonical ensemble of energy $E$.
The kinetic equations (\ref{eq:cin_eqs}), linearized around the stationary solution
$\{ \bar{n}_\qq(\bar{E}) \}$, can be put in the form
\be
\dot{\vec{x}}(\tau)=M \, \vec{x}(\tau)
\label{eq:lin_eqs}
\ee
where we have collected all the unknowns $n_\kk(\tau)-\bar{n}_\kk(\bar{E})$ in a single vector $\vec{x}(\tau)$ and $M$ is a matrix.
The existence of a zero frequency mode can be understood in two different ways that we explain.

\noindent
{\sl First reasoning}:  Consider an energy $E$ close to $\bar E$. In the same way as $\{ \bar{n}_\kk(\bar{E}) \}$, 
the set of occupation numbers $\{\bar{n}_\kk(E)\}$ constitutes a stationary solution of the full kinetic
equations (\ref{eq:cin_eqs}). Since the solutions are close, their difference $\{\bar{n}_\kk(E)-\bar{n}_\kk(\bar{E}) \}$
obeys the linear system (\ref{eq:lin_eqs}) so that the vector $\vec{e}_0$ of components
\be
e_{0 ,\kk} = \frac{d}{dE} \bar{n}_\kk(\bar{E})
\label{eq:e0k}
\ee
is a zero-frequency eigenmode of $M$.

\noindent
{\sl Second reasoning}: The Bogoliubov energy $\sum_{\kk \neq {\bf 0}}\epsilon_k n_\kk(\tau)$ is conserved by the kinetic
equations. An a consequence $\vec{\epsilon} \cdot \vec{x}(\tau)$ is a constant (the vector  $\vec{\epsilon}$ has
components $\epsilon_k$) and its time derivative is zero. This holds for all initial values of $\vec{x}$ , and thus implies
that $\vec{\epsilon}$ is a left eigenvector of $M$ with zero eigenvalue.
A basic theorem of linear algebra 
then implies the existence of a right eigenvector of $M$ with zero eigenvalue. Actually we already found it: it is
$\vec{e}_0$ of components (\ref{eq:e0k}). Such left and right eigenvectors are called adjoint vectors. For our normalization choice, 
their scalar product $\vec{\epsilon} \cdot \vec{e}_0=\frac{d}{dE}E=1$ as it should be.

We now go back to the correlation function $C(\tau)$. We introduce the (zero-mean) fluctuation operators
\be
\hat{\delta n}_\kk = \hat{n}_\kk - \bar{n}_\kk(\bar{E})
\ee
where we have neglected the difference between $\langle \hat{n}_\kk \rangle$ and $\bar{n}_\kk(\bar{E})$
in the large system size limit.
The correlation function $C(\tau)$ is then obtained as
\be
C(\tau)=\vec{A} \cdot \vec{x}(\tau) \hspace{0.5cm}\mbox{with}\hspace{0.5cm} 
x_\kk(\tau) = - \langle \delta \hat{n}_{{\bf k}}(\tau) \dot{\hat \theta}(0) \rangle
\ee
where we have collected in a vector $\vec{A}$, the coefficients in $\dot{\hat \theta}$ given by Eq.~(\ref{eq:thdot}):
\be
A_\kk \equiv \frac{g_0}{\hbar V} (U_k+V_k)^2
\ee
Following the reasoning of subsection \ref{sub:micro} on finds that $\vec{x}(\tau)$ obeys Eq.~(\ref{eq:lin_eqs}).
Splitting $\vec{x}(\tau) = \gamma \vec{e}_0 + \vec{X}\,(\tau)$ we have in the long time limit that 
$\vec{X}\,(\tau) \to 0$ due to the Beliaev-Landau damping processe whereas 
$\gamma=\vec{\epsilon} \cdot \vec{x}(0)$ is a constant.
At long times one then has
\be
C(\tau) \underset{\tau \to \infty}{\to} [\vec{\epsilon} \cdot \vec{x}(0)] (\vec{A} \cdot \vec{e}_0)
\ee
Taking the microcanonical average of  (\ref{eq:thdot}) and using (\ref{eq:thetadot_mc}) on obtains
the Bogoliubov expression for the microcanonical chemical potential:
\be
\mu_{\rm mc}(E) = \mu(T=0)(N) + \sum_{\kk \neq {\bf 0}}  \, \hbar \, A_\kk \bar{n}_\kk(E)
\ee
Using the  expression of $\vec{e}_0$ this leads to
$\vec{A} \cdot \vec{e}_0 = \frac{d}{dE}\mu_{\rm mc}({\bar E})/\hbar$.
We now evaluate the expectation value $\langle \ldots\rangle$ appearing in $\vec{\epsilon} \cdot \vec{x}(0)$ in two steps. 
We first take the expectation value in the microcanonical ensemble of energy $E$: one can then replace
the operator $\sum_\kk \epsilon_k \hat{\delta n}_\kk(0)$ with $E-\bar{E}$, since the total Bogoliubov energy is fixed
to $E$. One is left with a microcanonical average of $\dot{\hat{\theta}}(0)$ at energy $E$, an average already given
by Eq.~(\ref{eq:thetadot_mc}), and that one can expand around $\bar{E}$ to first order in $E-\bar{E}$. 
The last step is to average over $E$ with the probability distribution $P(E)$ defining the ensemble, to obtain
\be
\vec{\epsilon} \cdot \vec{x}(0) = \mbox{ Var}\,(E) 
"\frac{d\mu_{\rm mc}}{\hbar \, dE}(\bar{E})
\ee
Collecting all the results, we exactly recover the coefficient of $t^2$ in Eq.~(\ref{eq:gen_res}).

After this last reasoning, it becomes apparent  that, contrarily to the zero-frequency component $\gamma \vec{e}_0$,
 the contribution of the damped component $\vec{X}(\tau)$ of  $\vec{x}(\tau)$ can be treated to zeroth order in the energy
fluctuations: one can directly take $E=\bar{E}$ without getting a vanishing contribution to $C(\tau)$
and to Eq.~(\ref{eq:gen_res}). This explains why both the diffusion coefficient $D$ and the time offset $t_{\rm off}$, that
purely originate from $\vec{X}(\tau)$, are essentially ensemble independent.


\end{document}